\documentclass[letterpaper, 10 pt, conference]{ieeeconf}
\IEEEoverridecommandlockouts
\usepackage{amsfonts,amsmath,graphicx,mathenv}
\usepackage[hypertex,colorlinks=true]{hyperref} % pour dvi
\usepackage[]{authblk}

\newtheorem{thm}{Theorem}%
\newtheorem{rmk}{Remark}

\newcommand{\tr}[1]{\text{Tr}\left(#1\right)}

\newcommand{\RR}{{\mathbb R}}

\newcommand{\NN}{{\mathbb N}}

\newcommand{\dotex}{\frac{d}{dt}}
\newcommand{\ddotex}{\frac{d^2}{dt^2}}
\newcommand{\dddotex}{\frac{d^3}{dt^3}}

\title{\LARGE \bf Observer-based quantum state estimation by continuous weak measurement}
\author[1]{Zaki Leghtas}
\author[1]{Mazyar Mirrahimi}
\author[2]{Pierre Rouchon}

\affil[1]{\small{INRIA Paris-Rocquencourt
Domaine de Voluceau, BP105
78153 Le Chesnay Cedex, France}
}
\affil[2]{\small{Mines-ParisTech, Centre Automatique et Syst\`emes,
60, boulevard Saint-Michel
75272 Paris Cedex, France}}

\begin{document}

\bibliographystyle{IEEEtran}

\maketitle

\begin{abstract}
We propose to apply the Back and Forth Nudging (BFN) method used for geophysical data assimilations \cite{Auroux-Blum-NonLinProcGeophys_2008} to estimate the initial state of a quantum system. We consider a cloud of atoms interacting with a magnetic field while a single observable is being continuously measured over time using homodyne detection. The BFN method relies on designing an observer forward and backwards in time. The state of the BFN observer is continuously updated by the measured data and tends to converge to the systems state. The proposed estimator seems to be globally asymptotically convergent when the system is observable. A detailed convergence proof and simulations are given in the 2-level case. A discussion on the extension of the algorithm to the multilevel case is also presented.
\end{abstract}

\section{Introduction}
Estimating the state of a quantum system is a fundamental problem of great interest in quantum control. Amongst a variety of applications, it is essential to verify the efficiency of a quantum state preparation protocol \cite{Dotsenko-al-PRA_2009,Mirrahimi-VanHandel-PRA_2009}. For this reason, it is interesting to avoid the usual quantum state tomography scheme which involves doing the experiment many times and performing a strong projective measurement of a new observable at each preparation. Indeed, since many realizations of the preparation protocol are necessary to obtain one state estimation, the fidelity of the preparation protocol is averaged out over all these realizations. A new approach overcoming this problem was proposed and verified experimentally in \cite{Silberfarb-al-PRL_2005} where a controlled evolution is applied to an ensemble average while an observable is continuously measured. A Baysesian filter is then used to reconstruct the quantum state from the measurement record. In this paper, we consider a similar setting to the one in \cite{Silberfarb-al-PRL_2005,Smith-al-PRL_2004,Smith-al-PRL_2006}. We propose a new approach inspired of the BFN method used in geophysical data assimilation \cite{Auroux-Blum-NonLinProcGeophys_2008} to reconstruct the state of the system from the measured data. We design an observer, which is an estimation of the quantum systems state and feed in the data continuously until the observer converges to the systems state. A similar proposal was outlined in \cite{Donovan-Mirrahimi-Rouchon_2010}. However, the method we propose has the advantage of extending naturally to a multidimensional case and makes more use of the specific dynamics which the system undergoes. We guess this can strongly reduce the computation time of the estimation and increase it's robustness.\\
In section \ref{sec:problemsett} we detail the problem settings and give the dynamics equations of the BFN observer. Simulation plots are then presented in section \ref{sec:simulations} to demonstrate the efficiency of the state reconstruction protocol. A detailed convergence proof of the observer is given in section \ref{sec:convproof}. Finally, in section \ref{sec:multilevelcase}, we discuss the extension of this algorithm to the multilevel case.

\section{The problem setting}
\label{sec:problemsett}
We consider the experimental setting introduced in \cite{Silberfarb-al-PRL_2005}. To simplify the theoretical study, we suppose that the system is a spin $\frac{1}{2}$ system (instead of a system of total angular momentum equal to $3$ or $4$ as considered in \cite{Silberfarb-al-PRL_2005}). It interacts with a magnetic field in the x-y plane: the control, and a  probe. Homodyne detection of the probe, as explained in \cite{Smith-al-PRL_2004} enables a weak continuous measurement of the spin system. We suppose that all the parameters involved in the dynamics are known. The dynamics of the spin ensemble is described by the master equation:
\small
\begin{eqnarray}
\dotex{\rho(t)}&=&-i[B_x(t)\sigma_x+B_y(t)\sigma_y,\rho(t)]\notag\\
&+&\Gamma(\sigma_z\rho(t)\sigma_z-\rho(t))\label{eq:dynamics}\\
y(t)&=&\tr{\sigma_z\rho(t)}\notag
\end{eqnarray}
\normalsize
$[.,.]$ is the Lie Bracket operator and $\tr{.}$ is the trace operator. We take $\hbar=1$. $\rho(t)$ is the density matrix of the average ensemble at time $t$. It is a $2\times2$ positive Hermitian matrix of trace 1. $y$ is the measurement. $\sigma_x,\sigma_y$ and $\sigma_z$ are the standard Pauli matrices. $\Gamma>0$ gives the strength of the coupling between the probe and the system. The aim is, from a set of data $\{y(t)\backslash t\in[0,T]\}$, to estimate the initial state of the system $\rho(0)$ which can be any pure or mixed state.\\
In order to do so we use Luenberger observers based on the back and forth nudging (BFN) method \cite{Auroux-Blum-NonLinProcGeophys_2008,Donovan-Mirrahimi-Rouchon_2010}. The designed observer was first introduced, without BFN and for a non dissipative system, in \cite{Kosut-Rabitz-IFAC_2002}.\\
The idea is to design an observer on system \eqref{eq:dynamics} and another on the same system but by changing $t\rightarrow T-t$. And doing this iteration $n$ times. This is equivalent to supposing that the system has a periodic dynamics of period $2T$ which is symmetric with respect to $t=T$ and that we are measuring the system over a time interval $[0,2nT]$. This gives more time for the observer to converge with a small gain and with minimal amount of data. System \eqref{eq:dynamics} is referred to as the "forward" system and the same system changing $t$ to $T-t$ the "backward" system. The indice k introduced below refers to the $k^{th}$ back and forth iteration of the algorithm. The letter 'f' stands for "forward" and "b" for "backward". $\hat\rho$ is the designed observer.\\
For the forward system consider:
\small
\begin{eqnarray}
\dotex{\hat\rho^f_k(t)}&=&-i[B_x(t)\sigma_x+B_y(t)\sigma_y,\hat\rho^f_k(t)]\notag\\
&+&\Gamma(\sigma_z\hat\rho^f_k(t)\sigma_z-\hat\rho^f_k(t)) \notag\\
&-&(\Gamma+\gamma)\sigma_z(\hat y^f_k(t)-y(t) \label{eq:obs_forward} )\\
\hat y^f_k(t)&=&\tr{\sigma_z\hat\rho^f_k(t)} \notag
\end{eqnarray}
\normalsize

For the backward system consider:

\small
\begin{eqnarray}
\dotex{\hat\rho^b_k(t)}&=&i[B_x(T-t)\sigma_x+B_y(T-t)\sigma_y,\hat\rho^b_k(t)]\notag\\
&-&\Gamma(\sigma_z\hat\rho^b_k(t)\sigma_z-\hat\rho^b_k(t)) \notag\\
&+&(\Gamma-\gamma)\sigma_z(\hat y^b_k(t)-y(T-t))\label{eq:obs_backward} \\
\hat y^b_k(t)&=&\tr{\sigma_z\hat\rho^b_k(t)}\notag
\end{eqnarray}
\normalsize

Noting that $\hat\rho^b_k(0)=\hat\rho^f_k(T)$ and $\hat\rho^f_k(0)=\hat\rho^b_{k-1}(T)$ and $\gamma>0$.

We initialize the observer $\hat\rho^f_0(0)=\hat\rho(0)$ to be Hermitian and of trace 1. We typically take $\hat\rho(0)=\frac{1}{2}I$ where $I$ is the identity matrix. This way we make no a priori assumption on the initial state.

\begin{rmk}
Notice that the observer introduced above is trace preserving and stays Hermitian for all time. However it does not preserve positivity.
\end{rmk}

We define for all $t\in\RR^+$ ($k$ is defined as: $k=E(\frac{t}{2T})$ where $E$ represents the integer part)

\small
\begin{eqnarray}
\tilde\rho(t)&=&\hat\rho^f_k(t-2kT)-\rho(t-2kT)\notag\\
&&\qquad\qquad\text{  if  } t\in[2kT,(2k+1)T[\label{eq:tilderho}\\
&&\hat\rho^b_k(t-(2k+1)T)-\rho(2(k+1)T-t)\notag\\
&&\qquad\qquad\text{  if  } t\in[(2k+1)T,2(k+1)T[\notag\\
\tilde{B_x}(t)&=&B_x(t-2kT)\notag\\
&&\qquad\qquad\text{  if  } t\in[2kT,(2k+1)T[\notag\\
&&B_x(2(k+1)T-t)\notag\\
&&\qquad\qquad\text{  if  } t\in[(2k+1)T,2(k+1)T[\notag\\
\tilde{B_y}(t)&=&B_y(t-2kT)\notag\\
&&\qquad\qquad\text{  if  } t\in[2kT,(2k+1)T[\notag\\
&&B_y(2(k+1)T-t)\notag\\
&&\qquad\qquad\text{  if  } t\in[(2k+1)T,2(k+1)T[\notag\\
Z(t)&=& \tr{\sigma_z \tilde\rho(t)}
\end{eqnarray}
\normalsize

Let
$$
V: A \text{ (hermitian) } \rightarrow \tr{A^2}
$$
V is definite positive. For all $k\in\NN$ we have:

\small
\begin{eqnarray}
\dotex{V(\tilde\rho(t))}&=&-4\Gamma V(\tilde\rho(t))-2\gamma (Z(t))^2 \notag\\
&&\qquad \text{ if } t\in[2kT,(2k+1)T[\label{eq:dV1}\\
                        &=&4\Gamma V(\tilde\rho(t))-2\gamma (Z(t))^2 \notag\\
&&\qquad \text{ if } t\in[(2k+1)T,2(k+1)T[\label{eq:dV2}
\end{eqnarray}
\normalsize

We note, for all $k\in\NN$, $V_k=V(\tilde\rho(2kT))$. We define the function $g$ such that for all $t\in\RR^+$

\small
\begin{eqnarray*}
g(t)&=&e^{4\Gamma(t-2kT)} \text{  if  } t\in[2kT,(2k+1)T[\\
&&e^{-4\Gamma(t-2(k+1)T)} \text{  if  } t\in[(2k+1)T,2(k+1)T[
\end{eqnarray*}
\normalsize

\small
\begin{eqnarray}
V_{k+1}-V_k&=&-2\gamma (\int_{2kT}^{(2k+1)T}{g(t)Z^2(t)dt}\notag\\
\qquad\qquad&+&\int_{(2k+1)T}^{2(k+1)T}{g(t)Z^2(t)dt}) \label{eq:decreasingerror}\\
&\le& 0 \notag
\end{eqnarray}
\normalsize

$(V_k)_k$ is a decreasing sequence which is studied in more detail in Section \ref{sec:convproof}. Before looking into the convergence proof, we present some simulations which show the robustness of the convergence of $\hat\rho^f_k(0)$ towards $\rho(0)$ when $k$ goes to infinity.

\section{Simulations}
\label{sec:simulations}
\begin{figure}[!t]
\centering
\includegraphics[width=3in]{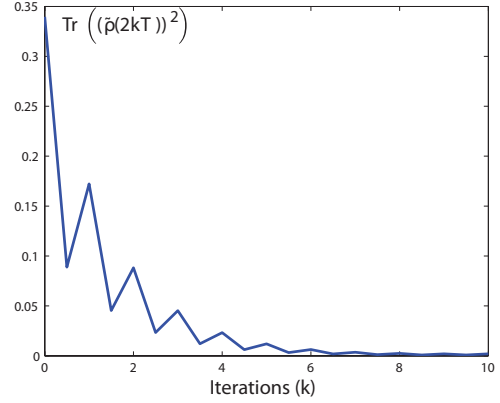}
\caption{Notice that the upper envelope (i.e $V_k=\tr{(\tilde\rho(2kT))^2}$) goes to zero when $k$ goes to infinity.}
\label{fig:lyapunov}
\end{figure}

\begin{figure}[!t]
\centering
\includegraphics[width=3in]{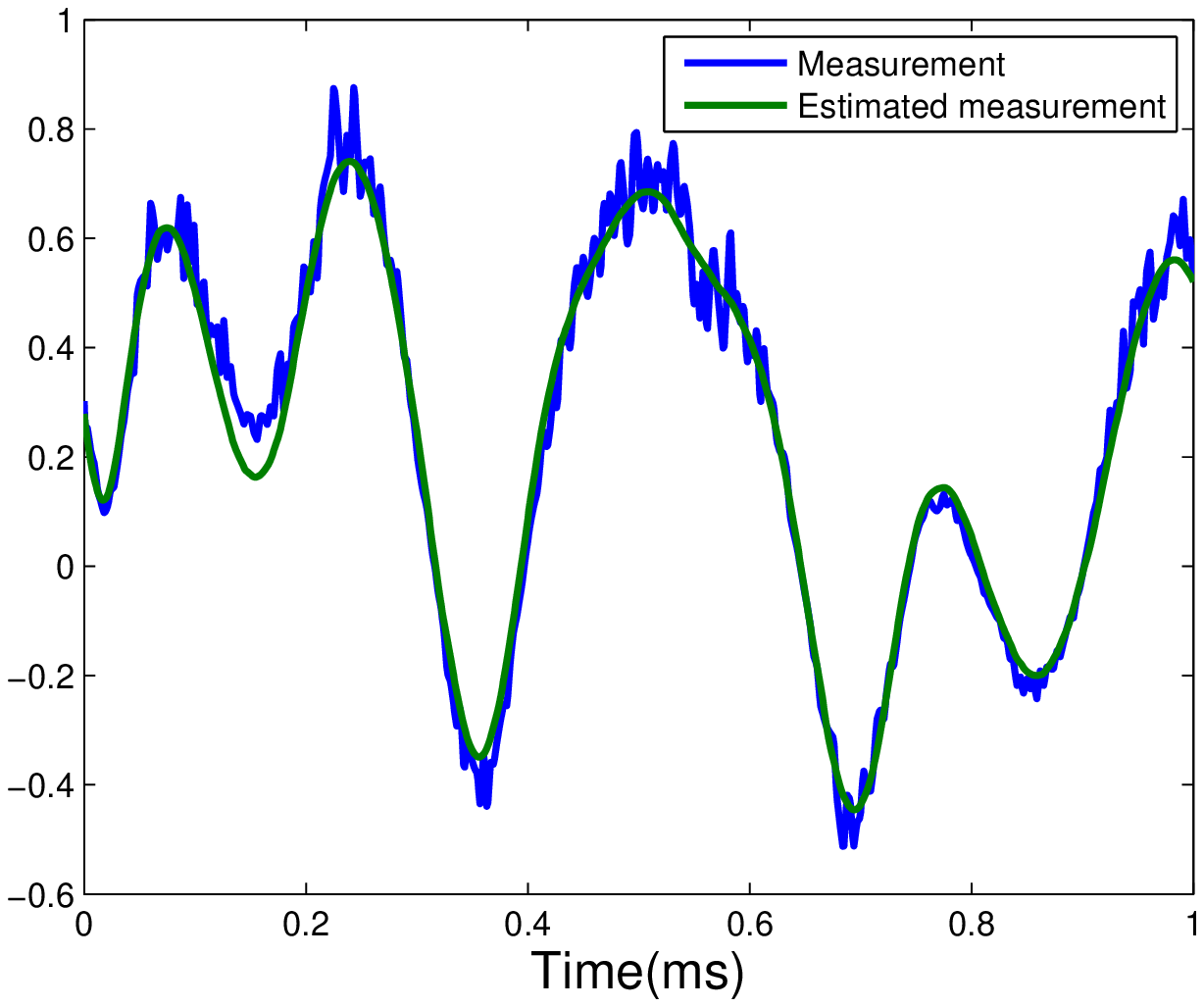}
\caption{Measurement and estimated measurement versus time. $10\%$ gaussian noise was added to the data. The estimated measurement is obtained by simulating \eqref{eq:dynamics} with $\rho(0)=\rho^f_k(0)$ and $k=10$.}
\label{fig:measurement}
\end{figure}

\begin{figure}[!t]
\centering
\includegraphics[width=3in]{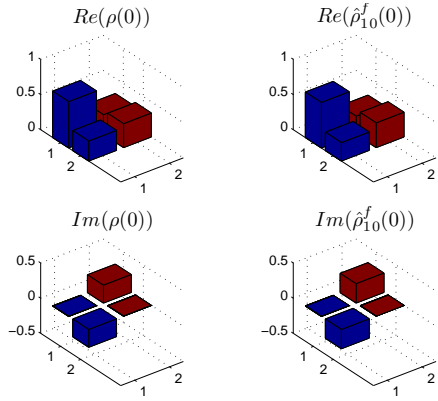}
\caption{The density matrix of the system and its estimator at time $t=0$.}
\label{fig:estimator}
\end{figure}

\begin{figure}[!t]
\centering
\includegraphics[width=3in]{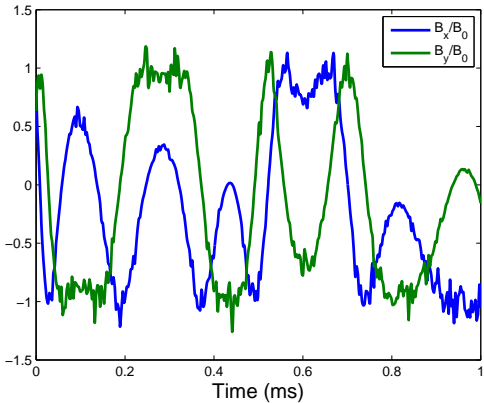}
\caption{The magnetic fields $B_x(t)$ and $B_y(t)$.}
\label{fig:fields}
\end{figure}

For the simulations of figures \ref{fig:lyapunov}, \ref{fig:measurement}, \ref{fig:estimator} and \ref{fig:fields}, we take:
\small
$$
\Gamma=0.25~ kHz \qquad \gamma=0.25~ kHz
$$
$$
\forall t ~~~ \sqrt{B_x(t)^2+B_y(t)^2}=10~ kHz \qquad T=1~ ms
$$
\normalsize
We take $B_x(t)=B_0\cos(\theta(t))$ and $B_y=B_0\sin(\theta(t))$. $B_0=10~ kHz$. For $\theta(t)$, at 10 equally spaced times between $0$ and $T$, we take a random value between $0$ and $2\pi$. Using a cubic spline interpolation, we build $\theta(t)$ over $[0,T]$.

10 iterations are simulated: $k=0,..,10$. $10\%$ gaussian noise was added to the measurement, $B_x$ and $B_y$.

We initialize the estimator in the completely mixed state $\hat{\rho}^f_0(0)=\hat\rho(0)=\frac{1}{2}I$. We randomly initialize $\rho$ on the Bloch sphere by taking a random Hermitian positive matrix satisfying: $\tr{\rho(0)^2}=\tr{\rho(0)}=1$.

\section{Convergence proof}
\label{sec:convproof}

\begin{thm}
\label{th:thm1}
For any $B_x,B_y \in \mathcal{C}^2([0,T],\RR)$ such that $B_x(0) \dotex{B_y}(0)-B_y(0) \dotex{B_x}(0) \ne 0$ and $\forall \hat\rho(0)$ which is Hermitian and of trace 1, we have
$$\lim_{t\rightarrow \infty} \tr{\tilde\rho^2(t)}=0$$
\end{thm}

\begin{rmk}
The convergence outlined in theorem \ref{th:thm1} is in two steps:
\begin{enumerate}
\item $\lim_{t\rightarrow \infty} Z(t)=0$
\item $\lim_{t\rightarrow \infty} \tr{\tilde\rho^2(t)}=0$
\end{enumerate}
The observer \eqref{eq:obs_forward}\eqref{eq:obs_backward} is designed such that we always have convergence of the estimated measurement to the measurement: $\lim_{t\rightarrow \infty} Z(t)=0$. Since the system is observable: $(\sigma_z,\sigma_x,\sigma_y)$ and its commutators span the space of all traceless $2\times2$ Hermitian matrices, we can find fields $(B_x,B_y)$ such that $\lim_{t\rightarrow \infty} Z(t)=0$ implies $\lim_{t\rightarrow \infty} \tr{\tilde\rho^2(t)}=0$.
\end{rmk}

\begin{proof}

For any piecewise continuous function $f$ we define: $f(2kT^+)=\lim_{t\rightarrow 2kT, t>2kT}f(t)$.

From \eqref{eq:decreasingerror}, we know that $(V_k)_k$ is a decreasing sequence. Besides, for all $k\in\NN$, $V_k \ge 0$, hence $(V_k)_k$ converges to a limit that we note by $V_\infty$. Summing \eqref{eq:decreasingerror} between $0$ and $N\in\NN^*$:

\begin{eqnarray*}
V_N-V_{0}&=&-2\gamma \int_{0}^{(2N+2)T}{g(t)Z^2(t)dt}\\
\end{eqnarray*}
$\forall t\in\RR^+$ $g(t)\ge 1$, hence $\int_{0}^{(2N+2)T}{Z^2(t)dt}\le \frac{V_0-V_{N}}{2\gamma}$. Since $\forall t\in\RR^+$ $Z^2(t)\ge0$, $\int_{0}^{\infty}{Z^2(t)dt}$ exists and is finite.\\

From \eqref{eq:dV1} and \eqref{eq:dV2} we have $\forall u\in[0,2T]$ and $\forall k\in \NN$:
\begin{equation}
V(\tilde\rho(2kT+u))\le V(\tilde\rho(2kT))\label{eq:Vinequality}
\end{equation}
hence, for all $t\in\RR^+$ we have $V(\tilde\rho(t))\le V_0$. $\tilde\rho$ is therefore bounded and belongs to the ball centered around $0$ and of radius $V_0$.\\
We are now going to prove that $(V_k)_k$ converges to zero when $k$ goes to infinity, and from \eqref{eq:Vinequality} we will conclude that $V(\tilde\rho(t))$ converges to zero when $t$ goes to infinity. In order to prove the convergence of $(V_k)_k$, we are going to prove that $Z(2kT),\dotex Z(2kT^+),\ddotex Z(2kT^+)$ all converge to zero when $k$ goes to infinity.

We consider $B_x,B_y$ $\mathcal{C}^2([0,T],\RR)$ functions. $Z$ is $\mathcal{C}^3$ over $S=\bigcup_{k\in\NN}]kT,(k+1)T[$.
$\tilde{B_x},\tilde{B_y},\dotex{\tilde{B_x}},\dotex{\tilde{B_y}},\ddotex{\tilde{B_x}},\ddotex{\tilde{B_y}}$ and $\tilde\rho$ are bounded over $S$, therefore $\dotex \tilde\rho,\ddotex\tilde\rho,\dddotex\tilde\rho$ are bounded over S and therefore $\dotex Z,\ddotex Z,\dddotex Z$ are bounded over S. Since $Z$ is continuous over $\RR^+$ and $\dotex Z$ is bounded over $S$, $Z$ is uniformly continuous over $\RR^+$, so $Z^2$ is uniformly continuous over $\RR^+$. What's more $\int_{0}^{\infty}{Z^2(t)dt}$ exists and is finite. We can conclude by applying Barbalat's lemma \cite{Khalil-NonLinSystems} that $\lim_{t\rightarrow \infty}Z^2(t)=0$ and hence
\begin{equation}
\label{eq:limZ}
\lim_{t\rightarrow \infty}Z(t)=0
\end{equation}
And in particular
\begin{equation}
\label{eq:limZkT}
\lim_{k\rightarrow \infty}Z(2kT)=0
\end{equation}

Since the derivatives of $Z$ are not continuous over $\RR^+$ but only over $S$, we cannot directly apply Barbalat's lemma to $\dotex Z(t)$ and $\ddotex Z(t)$.\\

Suppose that $\dotex{Z}(t)$ does not converge to zero when $t$ goes to infinity.\\
There exists $\epsilon>0$ and a sequence $(t_n)_n$ such that $\lim_{n\rightarrow \infty}t_n=\infty$ and $\dotex Z(t_n)>\epsilon$ (or $\dotex Z(t_n)<-\epsilon$ which can be treated in exactly the same way).\\
Since $\ddotex Z(t)$ is bounded over $S$ we have:
$\exists \eta\in]0,T/2[$ such that for all $n\in\NN$ and $t\in[-t_n^{min},t_n^{max}]$ $|\dotex Z(t_n+t)-\dotex Z(t_n)|<\frac{\epsilon}{2}$. Where $t_n^{min}=\min(t_n-E(t_n/T)T,\eta)$ and $t_n^{max}=\min((E(t_n/T)+1)T-t_n,\eta)$. $E$ represents the integer part.\\
Hence, for all $t\in[-t_n^{min},t_n^{max}]$ we have: $\dotex Z(t_n+t)=\dotex Z(t_n)-(\dotex Z(t_n)-\dotex Z(t_n+t))\ge\dotex Z(t_n)-|\dotex Z(t_n+t)-\dotex Z(t_n)|\ge \epsilon-\frac{\epsilon}{2}=\frac{\epsilon}{2}$. Also, notice that $T\ge t_n^{min}+t_n^{max}\ge \eta$.
\normalsize
\begin{eqnarray*}
&&|Z(t_n+t_n^{max})-Z(t_n-t_n^{min})|=\\
&&\qquad|\int_{t_n-t_n^{min}}^{t_n+t_n^{max}}{\dotex{Z}(t)dt}|\\
&&\qquad\ge\frac{\eta\epsilon}{2}\\
&&\qquad>0
\end{eqnarray*}
\normalsize

This is in contradiction with \eqref{eq:limZ}, we therefore conclude that
\begin{eqnarray}
\label{eq:limdotZ}
\lim_{t\rightarrow \infty}\dotex Z(t)&=&0
\end{eqnarray}
and in particular:

\begin{eqnarray}
\label{eq:limdotZkT}
\lim_{k\rightarrow \infty}\dotex Z(2kT^+)&=&0
\end{eqnarray}

Suppose that $\ddotex Z(2kT^+)$ does not converge to zero when $k$ goes to infinity, there exists $\epsilon>0$ and a sequence $(k_n)_n$ such that $\lim_{n\rightarrow \infty}k_n=\infty$ and $\ddotex Z(2k_nT^+)>\epsilon$, since $\dddotex Z$ is bounded over $S$, there exists $0<\eta<T$ such that for all $n\in\NN$ and $0<t<\eta$ $|\ddotex Z(2k_nT+t)-\ddotex Z(2k_nT^+)|<\frac{\epsilon}{2}$. Hence:

\begin{eqnarray*}
&&|\dotex Z(2k_nT+\eta)-\dotex Z(2k_nT^+)|=\\
&&\qquad|\int_{2k_nT}^{2k_nT+\eta}{\ddotex Z(t)dt}|\\
&&\qquad\ge\frac{\eta\epsilon}{2}\\
&&\qquad>0
\end{eqnarray*}

Which contradicts \eqref{eq:limdotZ}. Hence:

\begin{eqnarray}
\label{eq:limddotZkT}
\lim_{k\rightarrow \infty}\ddotex Z(2kT^+)&=&0
\end{eqnarray}

We note
\begin{eqnarray*}
X(t)&=&\tr{\sigma_x\tilde\rho(t)}\\
Y(t)&=&\tr{\sigma_y\tilde\rho(t)}
\end{eqnarray*}

We recall that from \eqref{eq:limZkT}\eqref{eq:limdotZkT}\eqref{eq:limddotZkT}:
\begin{System}
\label{sys:limitZs}
\lim_{k\rightarrow \infty} Z(2kT)=0\\
\lim_{k\rightarrow \infty}\dotex Z(2kT^+)=0\\
\lim_{k\rightarrow \infty}\ddotex Z(2kT^+)=0
\end{System}
Using \eqref{eq:dynamics}, we find that \eqref{sys:limitZs} implies:
\footnotesize
\begin{System}
\label{sys:lasalle}
\lim_{k\rightarrow \infty} Z(2kT)=0\\
\lim_{k\rightarrow \infty} \tilde{B_x}(2kT)Y(2kT)-\tilde{B_y}(2kT)X(2kT)=0\\
\lim_{k\rightarrow \infty} \dotex{\tilde{B_x}}(2kT)Y(2kT)-\dotex{\tilde{B_y}}(2kT)X(2kT)=0
\end{System}
\normalsize
Notice that $\tilde{B_x}(2kT)=B_x(0)$ and $\tilde{B_y}(2kT)=B_y(0)$, the same holds for their derivatives. We take $B_x,B_y$ such that $B_x(0) \dotex{B_y}(0)-B_y(0) \dotex{B_x}(0) \ne 0$.
\eqref{sys:lasalle} implies:

\begin{System*}
\lim_{k\rightarrow \infty} Z(2kT)=0\\
\lim_{k\rightarrow \infty} X(2kT)=0\\
\lim_{k\rightarrow \infty} Y(2kT)=0
\end{System*}

This is equivalent to $\lim_{k\rightarrow \infty}V(\tilde\rho(2kT))=0$. \eqref{eq:Vinequality} enables us to conclude that:

$$\lim_{t\rightarrow \infty}\tr{\tilde\rho^2(t)}=0$$

\end{proof}

\section{Extension to the multilevel case}
\label{sec:multilevelcase}
We now consider a system of total angular momentum $F$. The dimension of the system is $d=2F+1$, and the density matrix $\rho(0)$ belongs to the set of positive $d\times d$ Hermitian matrices of trace $1$. There are therefore $d^2-1$ parameters to identify. In order to extend the proof of the two level case ($d=2$) to the multilevel case, we would need to prove that $Z(t),\frac{d}{dt}Z(t),..,\frac{d^{d^2-2}}{dt^{d^2-2}}Z(t)$ all converge to zero when $t$ goes to infinity. If the system is observable, we would be able to conclude that we can find a control such that $\tilde\rho(t)$ converges to zero. Two complications arise from considering a system of higher dimension:\\
First, the need to extract information from further derivatives of the measurement record systematically reduces the robustness of the state estimation. One direction of improvement would be to design a nonlinear observer which preserves positivity. Such an observer is given in \cite{Leghtas-Mirrahimi-Rouchon-CDC_2009} for a non dissipative system ($\Gamma=0$). The difficulty is to build an observer which stays positive even in the backwards dynamics which is instable due to the dissipation term in $\Gamma$. Such an observer would reduce the size of the admissible $\rho(0)$s given a noisy measurement record, the robustness will therefore be increased. However, no dramatic improvement should be expected since the information on some matrix elements of $\rho(0)$ are hidden in high derivatives of the measurement record.\\
Second, although the observability criteria insures the existence of a control such that $\lim_{t\rightarrow\infty}Z(t)=0$ implies $\lim_{t\rightarrow\infty}\tilde\rho(t)=0$, we don't have any well known method to find such a control. The higher the dimension, the harder it is to find a control which makes the data $y(t)$ informationally complete about the initial state $\rho(0)$. \\
We now give some simulations which show that our BFN protocol still works well for a system of total angular momentum $F=1$ ($d=3$).
Consider the system:
\begin{eqnarray}
\dotex{\rho(t)}&=&-i[H(t),\rho(t)]+\Gamma\mathcal{D}[O]\rho(t)\label{eq:dynamicsmultilevel}\\
y(t)&=&\tr{O\rho(t)}\notag
\end{eqnarray}
Where $H$ is the systems Hamiltonian and $\mathcal{D}[O]$ the Lindblad superoperator. We have: $H(t)=g_F\mu_B(B_x(t)F_x+B_y(t)F_y)+\beta\Gamma {F_x}^2$. $g_F,\mu_B, \Gamma$ and $\beta$ are positive constants, $B_x,B_y$ are the controls and $F_x,F_y,F_z$ are the angular momentum operators. $O$ is the observable, and we take $O=\sqrt{\Gamma}F_z$. $\mathcal{D}[O]\rho(t)=O\rho(t)O^\dag-\frac{1}{2}(OO^\dag\rho(t)+\rho(t)O^\dag O)$. The superscript $^\dag$ stands for conjugate transpose. The term $\beta\Gamma {F_x}^2$ is necessary to insure the observability of the system \cite{Deutsch-Jessen-OpticsComm_2010}. We now consider the following observers:

\begin{eqnarray}
\dotex{\hat\rho^f_k(t)}&=&-i[H(t),\hat\rho^f_k(t)]+\Gamma\mathcal{D}[O]\hat\rho^f_k(t)\notag\\
&-&\gamma O(\hat y^f_k(t)-y(t))\label{eq:obs_forward_multi}\\
\hat y^f_k(t)&=&\tr{O\hat\rho^f_k(t)}\notag
\end{eqnarray}

\begin{eqnarray}
\dotex{\hat\rho^b_k(t)}&=&i[H(T-t),\hat\rho^b_k(t)]-\Gamma\mathcal{D}[O]\hat\rho^b_k(t)\notag\\
&-&\gamma O(\hat y^b_k(t)-y(T-t))\label{eq:obs_backward_multi}\\
\hat y^b_k(t)&=&\tr{O\hat\rho^b_k(t)}\notag
\end{eqnarray}

with the conditions: $\hat\rho^b_k(0)=\hat\rho^f_k(T)$ and $\hat\rho^f_k(0)=\hat\rho^b_{k-1}(T)$. We initialize the observer at $\hat\rho^f_0(0)=\frac{1}{3}Id$ where $Id$ is the $3\times 3$ identity matrix.

\begin{figure}[!t]
\centering
\includegraphics[width=3in]{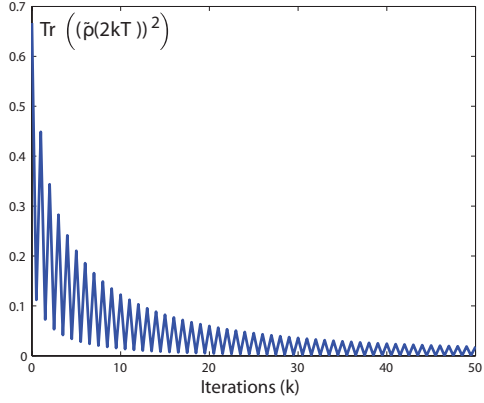}
\caption{The upper envelope is $V_k=\tr{(\tilde\rho(2kT))^2}$. Notice that it decreases and seems to converge to zero.}
\label{fig:lyapunov_multi}
\end{figure}

\begin{figure}[!t]
\centering
\includegraphics[width=3in]{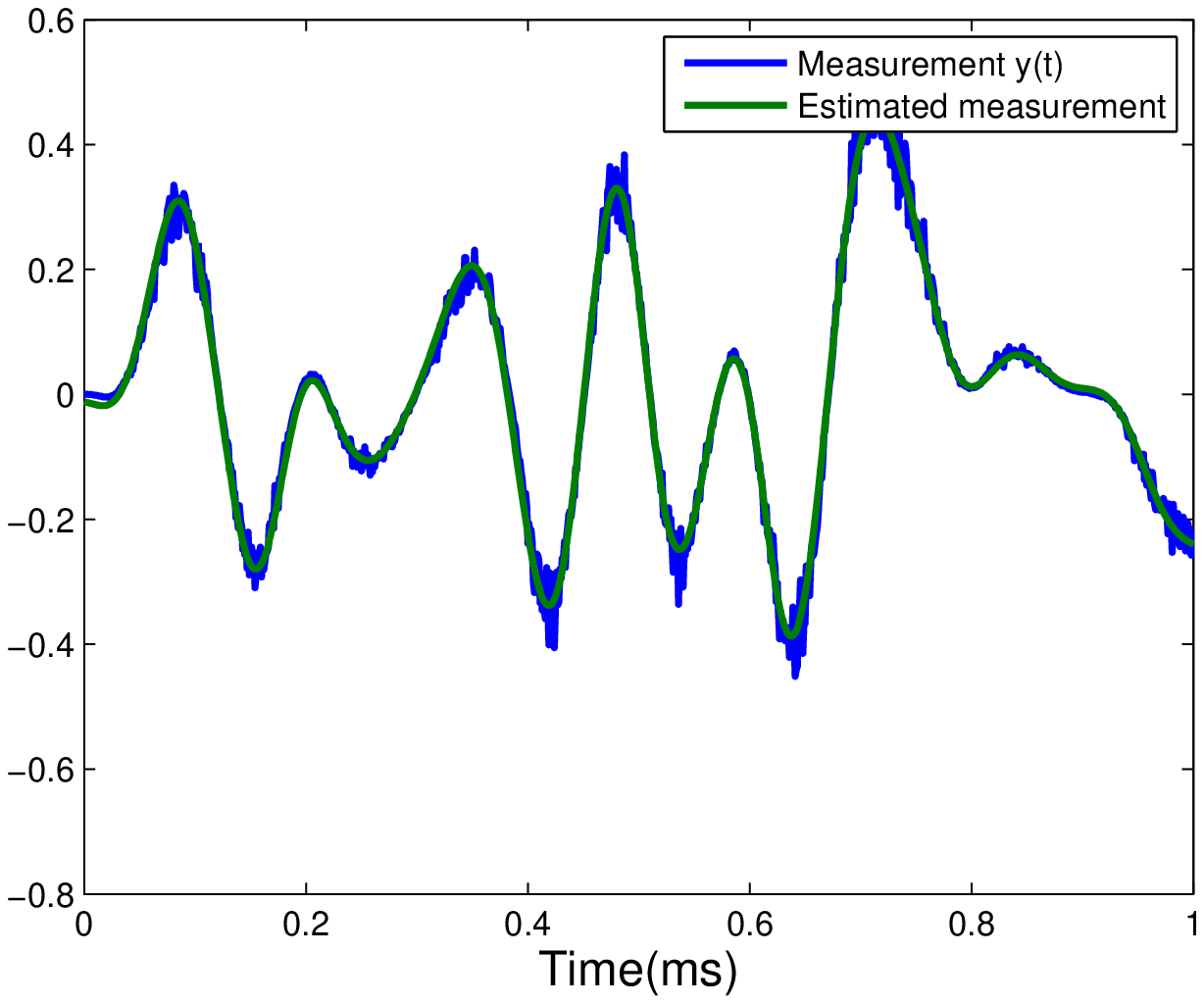}
\caption{Measurement and estimated measurement versus time. $10\%$ gaussian noise was added to the data. The estimated measurement is obtained by simulating \eqref{eq:dynamicsmultilevel} with $\rho(0)=\rho^f_k(0)$ and $k=50$.}
\label{fig:measurement_multi}
\end{figure}

\begin{figure}[!t]
\includegraphics[width=3in]{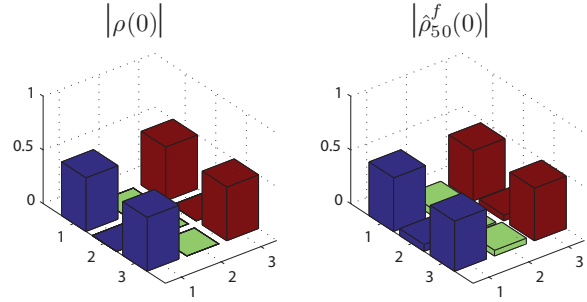}
\caption{The density matrix of the system and its estimator at time $t=0$. We plot the modulus of each matrix element.}
\label{fig:estimator_multi}
\end{figure}

For the numerical simulations in figures \ref{fig:lyapunov_multi}, \ref{fig:measurement_multi} and \ref{fig:estimator_multi} we take:
$$
g_F=1 \qquad \mu_BB_0=30 \qquad \Gamma=1 \qquad
$$
$$
\gamma=1 \qquad \beta=10 \qquad T=1
$$
We take $B_x(t)=B_0\cos(\theta(t))$ and $B_y(t)=B_0\sin(\theta(t))$. $\theta(t)$ is found using a numerical search routine aiming to maximize a certain criteria (entropy), as explained in \cite{Silberfarb-thesis_2006}. $10\%$ noise is added to the controls $B_x,B_y$ and $10\%$ noise is added to the measurement $y(t)$. We take
$\rho(0)=\frac{1}{2}
\begin{pmatrix}
1 & 0 & 1\\
0 & 0 & 0\\
1 & 0 & 1
\end{pmatrix}$

Notice that the estimated measurement is almost identical to the measurement $y(t)$ (figure \ref{fig:measurement_multi}). Also, the sequence $(V_k)_k$ decreases and seems to converge to zero (figure \ref{fig:lyapunov_multi}). This enables us to reconstruct the initial state with a $96\%$ fidelity where the fidelity $\mathcal{F}$ is computed as follows $\mathcal{F}=\tr{\sqrt{\sqrt{\hat\rho^f_{50}(0)}\rho(0)\sqrt{\hat\rho^f_{50}(0)}}}$ \cite{Jozsa-JModOpt_1994}. More iterations are needed than in the $2$ level case ($50$ as opposed to $10$) to achieve a similar fidelity. Each back and forth iteration takes about $0.1$ seconds so the presented simulation takes about $5$ seconds to run. As mentioned above, the fidelity of the reconstruction can be increased if a \emph{better} control is found.

\section{Conclusion}
In this paper we propose a BFN scheme to estimate the initial state of a quantum system when a continuous measurement of a single observable is given over a time interval $[0,T]$. A convergence proof and simulations are given for the two level case, and the considered experimental settings were similar to those in \cite{Smith-al-PRL_2004}. We discuss the extension of this algorithm to the multilevel case outlining the limitations and possible improvements of this protocol, and we present simulations in the case of a spin $1$ system. A quantitative comparison of this method to the ones considered in \cite{Silberfarb-al-PRL_2005} and \cite{Smith-al-PRL_2006} in terms of estimation time and robustness will be necessary to put forward the advantages of this state reconstruction protocol.

\section*{Acknowledgment}
This work was partly supported by the "Agence Nationale de la Recherche" (ANR), Projet Jeunes Chercheurs EPOQ2 number ANR-09-JCJC-0070

\bibliography{D:/Travail/bibliography}

\end{document}